# A Fast Pseudo-Stochastic Sequential Cipher Generator Based on RBMs

Fei Hu ·Xiaofei Xu ·Tao Peng ·Changjiu Pu ·Li Li[1]

**Abstract** Based on Restricted Boltzmann Machines (RBMs), an improved pseudo-stochastic sequential cipher generator is proposed. It is effective and efficient because of the two advantages: this generator includes a stochastic neural network that can perform the calculation in parallel, that is to say, all elements are calculated simultaneously; unlimited number of sequential ciphers can be generated simultaneously for multiple encryption schemas. The periodicity and the correlation of the output sequential ciphers meet the requirements for the design of encrypting sequential data. In the experiment, the generated sequential cipher is used to encrypt the image, and better performance is achieved in terms of the key space analysis, the correlation analysis, the sensitivity analysis and the differential attack. The experimental result is promising that could promote the development of image protection in computer security.

## 1 Introduction

In the field of communication and encryption for sequential data, sequential cipher generation algorithms have been one of the main technology used for military and diplomatic occasions. As a kind of symmetric encryption algorithm, sequential ciphers have the following characteristics: easy to be implemented, simple implementation with hardware, fast in encryption and decryption processing, none or limited error propagation. Shannon proved the one-time pad encryption system was safe [1]. It plays an important role in promoting the development of the sequential ciphering technology. The development of the sequential cipher technology has been attempting to imitate the one-time pad scheme, i.e., the one-time pad encryption system is the prototype of the sequential ciphering system. In order to cipher the sequential data, a stochastic sequence, determined by a cipher code, will be generated at first. The algorithms of generating stochastic sequential ciphers can be roughly divided into two categories: Linear Feedback Shift Register sequential cipher generators (LFSR) and nonlinear sequential cipher generators. LFSR

Fei Hu

School of Computer and Information Science, Southwest University, Chongqing, China

Network Centre, Chongqing University of Education, Chongqing, China

Xiaofei Xu

School of Computer and Information Science, Southwest University, Chongqing, China

Tao Peng

Network Centre, Chongqing University of Education, Chongqing, China

Changjiu Pu

Network Centre, Chongqing University of Education, Chongqing, China

Li Li

School of Computer and Information Science, Southwest University, Chongqing, China

[1] Corresponding author. E-mail: lily@swu.edu.cn

algorithms generate sequential ciphers by maximizing the length of the shift registers and by using a linear feedback function. The theoretical foundation of LFSR algorithms has been mature [2–4]. Sequential ciphers generated by LFSR algorithms are stochastic, however, the resulting sequential ciphers have the risk of being decrypted [5]. As a result, people turn to the nonlinear field instead.

The encryption algorithms using nonlinear sequential ciphering technology can be roughly divided in to the following categories: (1) the Nonlinear Feedback Shift Register sequential cipher generator (NFSR), it consists of a shift register and a nonlinear feedback function, and can yields a 2n-length sequence, NFSR has good cryptographic properties, but the generation speed is slow and restricts the development of NFSRs; (2) the nonlinear-combination sequential cipher generator, it consists of multiple LFSRs and nonlinear functions; (3) the clock controlled sequential cipher generator, it controls registers using another registers clock, it is easy to be attacked using logistic analysis; (4) other sequential cipher generators, these generators construct complex networks to generate sequential ciphers by combining above algorithms, e.g. the chaotic theory, the artificial neural network, the DNA encoding technology and the quantum encryption.

Probably Lauria was the first to use artificial neural networks (ANNs) in cryptography [6–8]. ANN has the following characteristics that are suitable for cryptography: Nonlinear calculation, associative memory, massively parallel processing and strong fault tolerance. With these characteristics, cryptography has a broad application prospect in the field of Very Large Scale Integration (VLSI) and optical implementations. In [9], Ding et. al. used a discrete Hopfield neural network to make a nonlinear sequential cipher generator. The network can generate sequential ciphers meeting the requirements for the design of encrypting sequential data. By using a small amount of stochastic parameter as the cipher codes, pseudo-stochastic sequential ciphers can be generated, they are desirable for information encryption. In [10–12], for example, ANN models were used for image encryption. Among all kinds of ANN models, the Boltzmann Machine (BM) [13,14] model has a good effect in generation of stochastic sequences. BM was first proposed by Hinton et. al. in 1986. It is a stochastic neural network rooted in the statistical mechanics with two output states for each neuron: activated or inactivated, commonly expressed as binary 1 or 0. The two states are taken according to the decision rule by probability: If the neuron has a higher activation probability, its state is 1; If otherwise, its state is set to 0.

The BM is a feedback neural network and neurons are fully connected without self-connection. Using Hopfield networks, it is easy to fall into the problem of a local minimum, while using the BM, the problem is well solved. In the actual learning process, some stochastic variables are unobservable (they are called hidden variables). By having more hidden variables, the modeling capacity of the BM is increased. The Restricted Boltzmann Machine (RBM) [15, 16] further restrict the BM by dividing it into two layers: a visible layer and a hidden layer. Variables in the visible layer are conditional independent, and so do variables in the hidden layer. Variables between the visible layer and the hidden layer are connected. By maximizing the likelihood function, RBMs optimize parameters and learn the probability distribution of the observed data. The parameters learned by RBMs meet the requirements of the design of stochastic sequences, and can be used to encrypt sequential data.

In this paper, a pseudo-stochastic sequence generation algorithm was proposed based on the RBM. For the convenience of handling images, the values of visible and hidden layers are defined as the floating-point numbers, and the pixel values of images are normalized between 0 and 1. The generated sequence was then used to encrypt an image. The results of the correlation analysis and sensitivity analysis proved that the sequence completely meets the requirements of image encryption. The algorithm can be used for real-time image transmission and protection on the internet. Compared with conventional pseudo-stochastic sequence generation algorithms, such as the logistic map, the algorithm in this paper is more powerful. In other words, the model is able to calculate all elements of a sequence in a single cycle calculation rather than one element in a cycle, and can generate a large number of required sequences simultaneously.

## 2 Restricted Boltzmann Machines

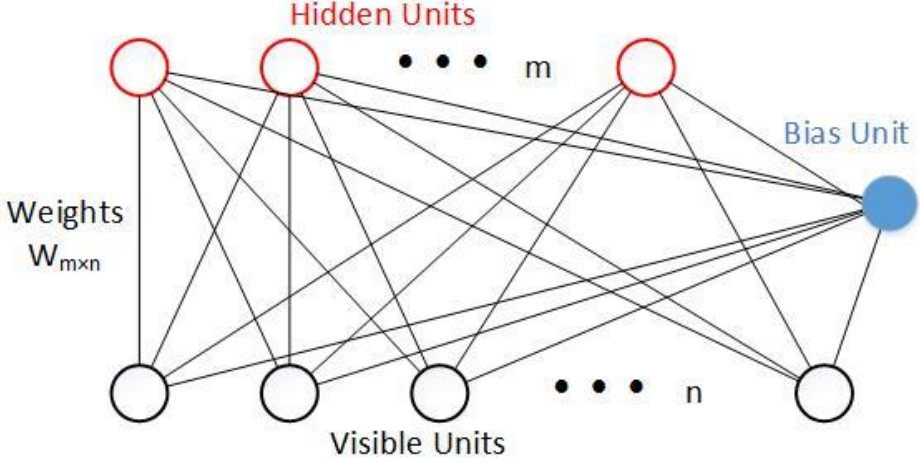

Fig. 1: Restricted Boltzmann Machines (RBM)

The RBM model is a bipartite-graph model, see Fig. 1. Each stochastic variable is represented by a node, marked as a cycle, also called a neuron. The interdependence between any pair of variables is represented by a line. Neurons have three types: the black cycles below represent observed variables, called visible variables, they make the visible layer, which commonly is used as the interface for input data; the red cycles above represent hidden variables, which meet the conditional probability distribution given the visible variables; the solid blue cycle on the right side represents the offset value for adjusting the values of learning parameters. Any two Neurons from different types are correlated, yet any two neurons from the same type are uncorrelated. The above RBM network structure has n visible neurons and m hidden neurons. The state of each visible neuron is only affected by the m hidden neurons, and the state of each hidden neuron is just affected by the n visible neurons. This characteristic makes the RBM model easy to train. The RBM network has two sets of parameters, also called weights, one is the weight matrix between the visible and hidden layers, and the other one is the set of offset values for the two layers. The two sets of parameters determine what kind of m-dimensional output data is encoded from an n-dimensional sample. The probability distributions meet the following equations:

$$p(h_j = 1|V) = \delta(\sum_{i=1}^{n} w_{ji} \times v_i + b_j) \quad (1)$$
$$p(v_i = 1|H) = \delta(\sum_{j=1}^{m} w_{ji}^T \times h_j + b_i) \quad (2)$$

Equation (1) calculates the conditional probability of the hidden variables given the visible variables; equation (2) calculates the conditional probability of the visible variables given the hidden variables. Where V is the set of the visible variables, H is the set of the hidden variables, b is the set of the offset variables, $\delta(\cdot)$ is a nonlinear function. A random number r is predefined as a threshold which has a value between [0,1]. When $p(h_j = 1|V) \geq r$, it is supposed that the probability of $h_j = 1$ is big enough, then $h_j = 1$; on the contrary, it is supposed that the probability of $h_j = 1$ is small, then $h_j = 0$. $p(v_i = 1|h)$ is also applicable to the above definition.

RBMs are a special form of log-linear Markov Random Field (MRF), i.e., for which the energy function is linear with its free parameters. By the introduction of the hidden neurons, the expression ability of the model is enhanced, which can indicate any complex probability distribution without prior knowledge of the probability distribution of the visible variables. And the introduction of the energy function makes it easier to train the model. Using equations (1) and (2), the model is trained to minimize the energy function with the gradient descent algorithm, and ultimately to maximize the fitting of the input data and the output data. In the training process, the weight matrix W is randomly disturbed. The sequence extracted from the disturbed W meets the requirements of the design of stochastic sequential ciphers. The energy function of RBM is as following:

$$E(v, h) = -\sum_{i=1}^{n}\sum_{j=1}^{m} w_{ij} v_i h_j - \sum_{i=1}^{n} b_i v_i - \sum_{j=1}^{m} b_j h_j \quad (3)$$

where $w_{ij}$ is the connection weight between the visible and hidden layers, $b_i$ is the offset value of the visible layer, $b_j$ is the offset value of the hidden layer. All neurons together have one energy value, i.e., any set of values for A Fast Pseudo-Stochastic Sequential Cipher Generator Based on RBMs all neurons have an energy value. For example, the values of neurons in the visible layer are (1,0,1,0), the values in the hidden layer are (1,1,0), then an energy is got by all the values putting into the equation (3). The energy of all visible neurons and hidden neurons is accumulated, and the result is set as the objective function of the RBM model. But the objective function is hard to solve, that is, through the RBM model, each input sample has a large amount of output values, the more the number of visible and hidden neurons, the larger the amount of values, each pair of input sample and output value has an energy, such a large amount of pairs bring exponential levels of calculation. So the solution is not practical, especially by using the exhaustive method to calculate the gradient descent.

In order to solve the above problems for the energy model, a joint probability function for visible and hidden neurons is introduced following:

$$p(v, h) = \frac{e^{-E(v,h)}}{\sum_{v,h} e^{-E(v,h)}} \quad (4)$$

The joint probability function p (v, h) is determined by the energy of the visible neurons and the hidden neurons. This function is defined based on the energy definition of statistical thermodynamics: When the system is in a thermal equilibrium with the surrounding environment, the probability of state i is as follows:

$$p_i = \frac{1}{Z} \times e^{-\frac{E_i}{k_b \times T}}$$

$$Z = \sum_{v,h} e^{-E(v,h)} \quad (5)$$

where $E_i$ is the energy of the system in the state i, T is the Kelvin absolute temperature value, k is a Boltzmann constant, and Z is a state-independent constant. In the equation (4), $E_i$ is turned into E (v, h), T and k are set as 1, Z is the denominator that makes sure the sum of all probabilities is 1.

The probability distribution of the observed data is derived from equation (4):

$$p(v) = \frac{\sum_h e^{-E(v,h)}}{Z} \quad (6)$$

In statistical mechanics, it is considered that the state with lower energy has a higher probability. The process of solving such a statistical mechanics model is to find a state with the lowest energy, when the probability of the state has the highest value. The following equation is Free Energy function, which is from statistical mechanics.

$$\text{FreeEnergy}(v) = -\ln \sum_h e^{-E(v,h)} \quad (7)$$

Then p (v) can be written in the following form:

$$p(v) = \frac{e^{-FreeEnergy(v)}}{Z} \quad (8)$$

After logarithm fetch on both sides of equation (8), which gets the following equation:
$$lnp(v) = -FreeEnergy(v) - lnZ \quad (9)$$
The Free Energy can be measured by ln(p(v)), i.e., p(v) is large when the Free Energy is small. After accumulation on both sides of equation (9), which gets the following equation:
$$\sum_v \ln p(v) = -\sum_v FreeEnergy(v) - \sum_v lnZ \quad (10)$$
Z is the sum of total energy between visible and hidden layers. And the computational complexity is $O(2^{m+n})$. It is impossible to do such a huge computation. The Contrastive Divergence (CD-k) was introduced by Hinton et al. to approximate it [17]. Substituting equation (7) into (10), we get the following equation:
$$\sum_v \ln p(v) = \sum_v \ln \sum_h e^{-E(v,h)} - \sum_v lnZ \quad (11)$$
The above equation is the negative value of the sum of total Free Energy of the whole network, that is, the RBM model. Because the equation is the logarithm sum of all probabilities p (v), the equation can be referred to as the log-likelihood function.

When the maximum likelihood estimate reaches the maximum, the energy value of the system is minimum. Because RBM is a random machine, it is dependent on the probability to evaluate its performance. The criterion of the evaluation is the likelihood function, i.e., the maximum likelihood estimation is used to train the network and to find the right parameters. Which makes sure the training samples have the maximum probability using the network with the right parameters. $\sum_v \ln p(v)$ is the log-likelihood function.

**3 Pseudo-Stochastic Sequential Cipher Generation Algorithm**

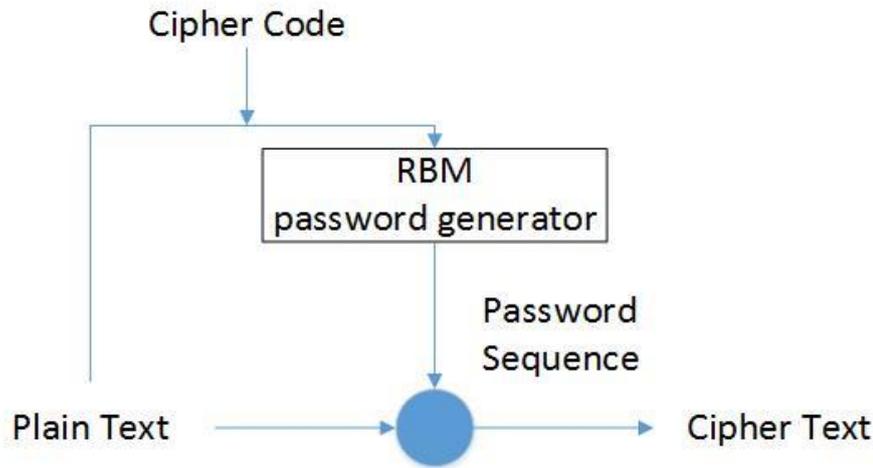

Fig. 2: Encryption device.

The diagram in Fig. 2 shows the algorithm of sequential cipher generation, it is described as follows:
1) Initialization:
   a cipher code X is predefined, has a value between 0 and 1. X is taken as a seed to initialize the weight parameters of the RBM model. The iteration number of training is N. The training will be ceased when the iteration number reaches N. The plain text is converted to one-dimensional vector v, the length of v is n, and the number of neurons in the visible layer of the RBM model is set as n. To set the number of neurons in the hidden layer as arbitrary m. For the convenience of handling images, we modified the RBM model, redefined the visible and hidden layers as float-point numbers. The pixel values of images are normalized between 0 and 1 as the input of the visible layer;

2) Training the RBM model:

the activation function σ(·) is a nonlinear function, the sigmoid function is used in the experiment. The likelihood in equation (11) is maximized by adjusting the weights using the Contrastive Divergence (CD-k) [17];

3) Iteration:

step 2 is repeated until the iteration number reaches N;

4) Pseudo-stochastic sequential cipher:

the vector $W_{in}$ is a Pseudo-stochastic sequential cipher, where i ∈ 1, 2, ...., m, n is the length of v;

5) Encryption and decryption:

the encryption and decryption functions are described following, where C is the cipher text, v is the plain text, bitxor( ) is an XOR function.

$$C = bitxor(V, W_{in}) \quad (12)$$
$$V = bitxor(C, W_{in}) \quad (13)$$

The practicability of the algorithm will be verified in the next section through image encryption.

# 4 Experiments

## 4.1 Experimental settings

This encryption algorithm was evaluated on four randomly selected gray-level images from the standard set of images which are frequently used in many literatures [18–20]. They are Lena, Lake, Yacht and Baboon. These images have the same size of 512*512 pixels. Each pixels component has 256 gray levels. That is an integer between 1 and 256. So every image is represented by a 512*512 digital matrix. The matrix was normalized, i.e., the matrix was divided by 256 and each element had a digital value between 0 and 1. The normalized matrix was shaped into a one-dimensional vector v. The iteration number N=100. v and a predefined cipher code X were put into the algorithm described in section 3, then m pseudo-stochastic sequential ciphers were got. One of the ciphers are put into the equations (12) and (13), the image was encrypted and the encrypted image was decrypted separately. The original images, the corresponding encrypted images and decrypted images are listed in Table 1. Every original image and its corresponding decrypted image are almost indistinguishable. In the next subsection, several criteria will be used to evaluate the quality of this algorithm, and comparative study will be performed with other algorithms.

## 4.2 Performance and security analysis

1) Histogram analysis.

An image can be read because it has a wealth of statistical information. Which reflects the frequency of each gray level in the image. The histograms in Table 2 tell that correlations between pixels in the encrypted image are greatly minimized. So the effect of image encryption is good.

2) Key space analysis.

The cipher code and parameters relative with the stochastic sequential ciphers can be described as: key= {X, Y, m}, where X is the calculation precision in MATLAB, its default value is $10^{16}$. Y is the iteration number of training on the RBM model, m is the number of neurons in the hidden layer of the RBM model. In this experiment, Y is $10^2$, m is 512*512 which is approximately $10^5$. So the key space of the encryption schema is $10^{23}$, and the key space can be larger if the three parameters of the key space become bigger. It is extremely hard to decrypt within such a large key space.

*Table 1.* Image encryption and decryption.

| | Original images | Encrypted images | Decrypted images |
|---|---|---|---|
| Lena | 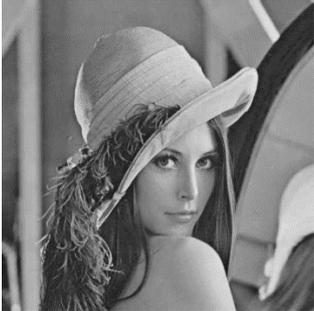 | 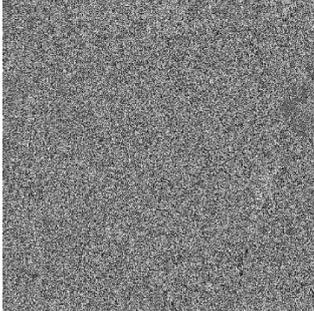 | 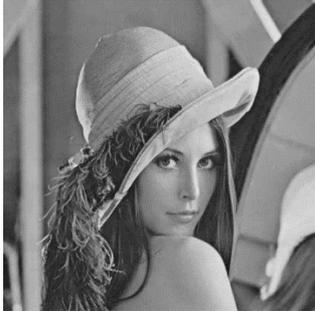 |
| Lake | 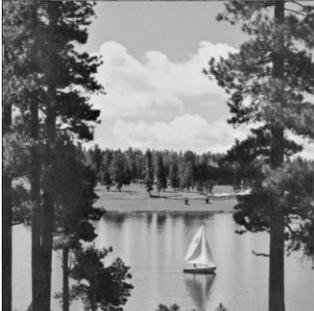 | 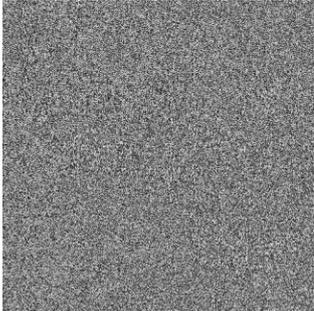 | 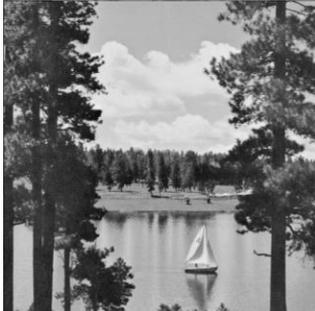 |
| Yacht | 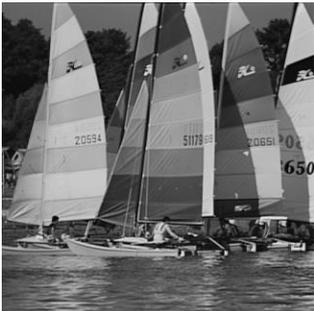 | 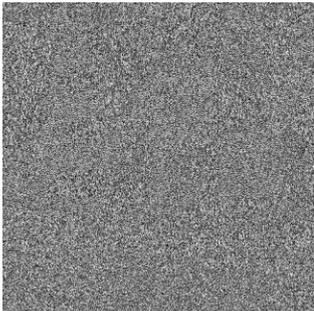 | 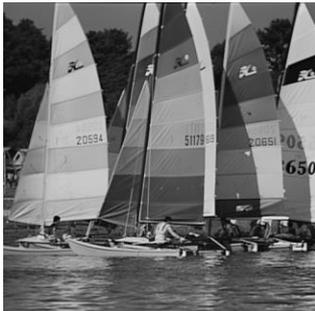 |
| Baboon | 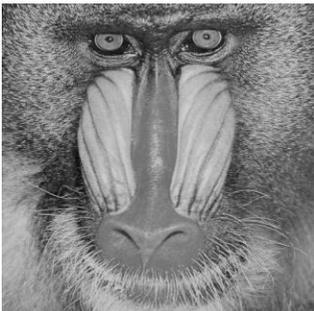 | 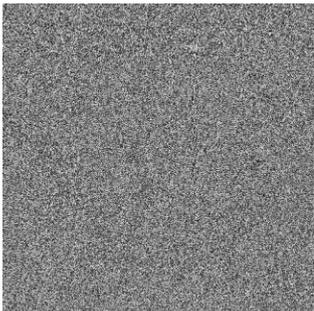 | 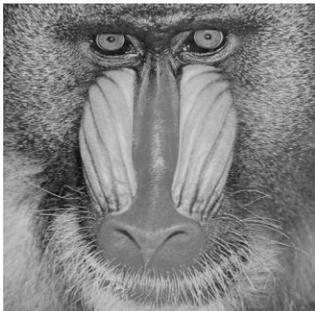 |

3) Correlation analysis.

A rich semantic image has high correlations among adjacent pixels. An encryption schema decreases the correlations between adjacent pixels in an image. The less correlation between pixels, the more difficult to perform cracking. The correlation coefficient is used to evaluate the correlation of a pair of adjacent pixels, and it is defined below [21]:

$$r_{xy} = \frac{cov(x,y)}{\sqrt{D(X)}\sqrt{D(y)}} \qquad (14)$$

$$E = \frac{1}{N}\sum_{i=1}^{N} x_i \quad (15)$$

$$D(x) = \frac{1}{N}\sum_{i=1} N(x_i - E(x))^2 \quad (16)$$

$$Cov(x, y) = \frac{1}{N}\sum_{i=1}^{N}(x_i - E(x))(y_i - E(y)) \quad (17)$$

where $r_{xy}$ is the correlation coefficient of the variables x and y, E ( ) is the mean function, D ( ) is the variance function, Cov ( ) is the covariance function, x and y are adjacent pixels.

*Table 2.* Histograms of original and encrypted images.

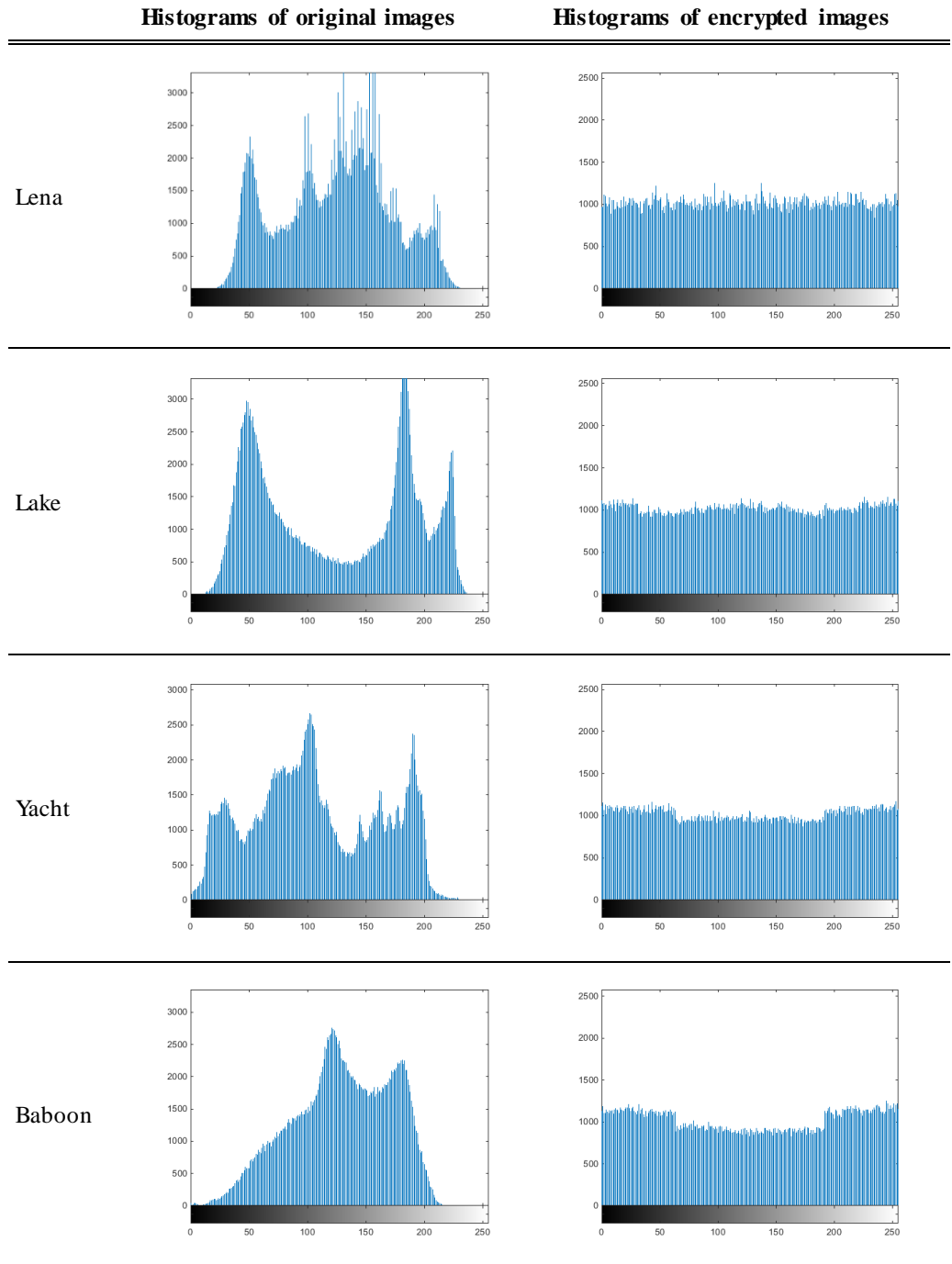

Ten pairs of adjacent pixels were selected in every original image and the corresponding encrypted image separately. And the correlation coefficients of these pairs are shown in Table 3. The smaller the correlation coefficient, the better the encryption effect. The original image contains rich semantic information, so adjacent pixels of it have strong correlation; yet adjacent pixels of the encrypted image have small correlation coefficients, almost close to zero, so they have a weak correlation. This table proves that the encryption algorithm is feasible.

*Table 3.* Correlation analysis.

| Images | Correlation coefficients | | | | |
|---|---|---|---|---|---|
| Original Lena | 0.9842 | 0.9828 | 0.9866 | 0.9825 | 0.9816 |
| | 0.9830 | 0.9869 | 0.9868 | 0.9845 | 0.9837 |
| Original Lake | 0.9886 | 0.9857 | 0.9835 | 0.9855 | 0.9842 |
| | 0.9866 | 0.9825 | 0.9866 | 0.9837 | 0.9868 |
| Original Yacht | 0.9834 | 0.9855 | 0.9825 | 0.9846 | 0.9862 |
| | 0.9839 | 0.9842 | 0.9857 | 0.9838 | 0.9844 |
| Original Baboon | 0.9826 | 0.9815 | 0.9844 | 0.9838 | 0.9857 |
| | 0.9846 | 0.9833 | 0.9828 | 0.9826 | 0.9843 |
| Encrypted Lena | 0.0067 | 0.0074 | -0.0082 | -0.0099 | -0.0191 |
| | -0.0125 | 0.0230 | 0.0479 | 0.0524 | -0.0307 |
| Encrypted Lake | 0.0132 | 0.0071 | -0.0270 | 0.0052 | 0.0186 |
| | 0.0421 | 0.0151 | 0.0023 | -0.0523 | 0.0067 |
| Encrypted Yacht | 0.0017 | 0.0053 | 0.0471 | -0.0036 | 0.0270 |
| | -0.0189 | 0.0037 | -0.0097 | 0.0241 | -0.0059 |
| Encrypted Baboon | -0.0230 | 0.0157 | 0.0482 | 0.0097 | 0.0241 |
| | -0.0082 | 0.0163 | 0.0367 | -0.0230 | -0.0152 |

The correlation distribution in an image is usually presented by a scattered diagram. We randomly selected 1000 pairs of pixels in each image either in horizontal, vertical, or diagonal direction, and plotted the scatter diagram, respectively. These diagrams are listed in Table 4. There is a strong correlation for each pair of adjacent pixels of the original image in three directions separately, but the encrypted image conceals the significant characteristics in any of the three directions.

4) Sensitivity analysis of the cipher code.

Even if the cipher code has a small change, the encrypted image cannot be decrypted. That is, the image encryption algorithm is robust.

The cipher code X was increased by a very small number, say $10^{-15}$ in this experiment. We then obtain a new cipher code X' which is equivalent to the original cipher code plus this $10^{-15}$. The encrypted images by the original cipher code X were decrypted using the changed cipher code X', where the encrypted images were listed in Table 1 and the decrypted images were shown in Fig. 3. It shows that every decrypted image cannot be recognized even partially with such a tiny change in the cipher code. That is to say, the algorithm is very sensitive and any random guess did not work at all. It thus guarantees all the encrypted images.

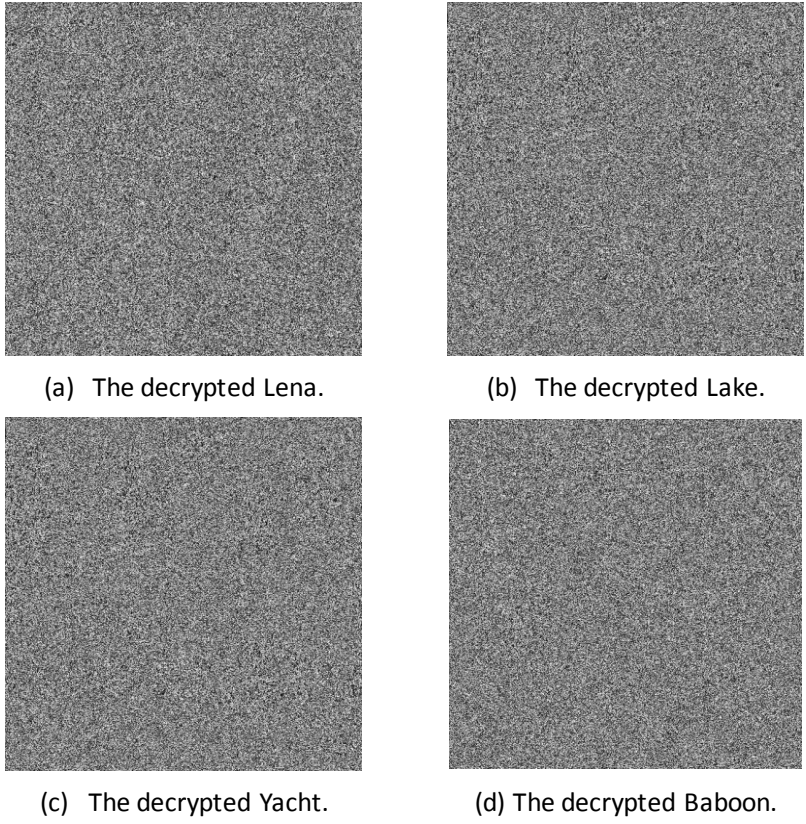

(a) The decrypted Lena.  (b) The decrypted Lake.

(c) The decrypted Yacht.  (d) The decrypted Baboon.

Fig. 3 The decrypted images with a changed cipher code.

5) Differential attack.

The encrypted image should be significantly different when its original one makes a small change. Otherwise the encryption schema would be vulnerable from differential attack. Number of pixels change rate (NPCR) and unified average changing intensity (UACI) can be used to measure this capability. NPCR reflects the degree of change of two encrypted images whose original images have only one different pixel. UACI denotes the average intensity of the change of two encrypted images with one pixel different in their original images. The NPCR and UACI are defined as follows [22,23]:

$$NPCR = \frac{\sum_{i,j} D(i,j)}{W \times H} \times 100\% \quad (18)$$

$$UACI = \frac{1}{W \times H} \left[ \sum_{i,j} \frac{|E_1(i,j) - E_2(i,j)|}{255} \right] \times 100\% \quad (19)$$

where $E_1$ and $E_2$ are the two encrypted images with one pixel different in their original images, $E_1(i, j)$ and $E_2(i, j)$ are the gray-scale pixel values at position (i, j) of the two encrypted images $E_1$ and $E_2$ respectively, W is image width, H is image height, D (i, j) is defined as:

$$D(i,j) = \begin{cases} 1, & if\ E_1(i,j) \neq E_2(i,j) \\ 0, & if\ E_1(i,j) = E_2(i,j) \end{cases} \quad (20)$$

The values of the two measures (NPCR and UACI) for our algorithm are listed and are compared with other encryption schemas in Table 5. The NPCR is over 99% and the UACI is over 33%. Which shows that our algorithm is very sensitive with respect to little change in the original image. The comparative results show that our algorithm is effective.

*Table 4.* Correlation distribution of two adjacent pixels.

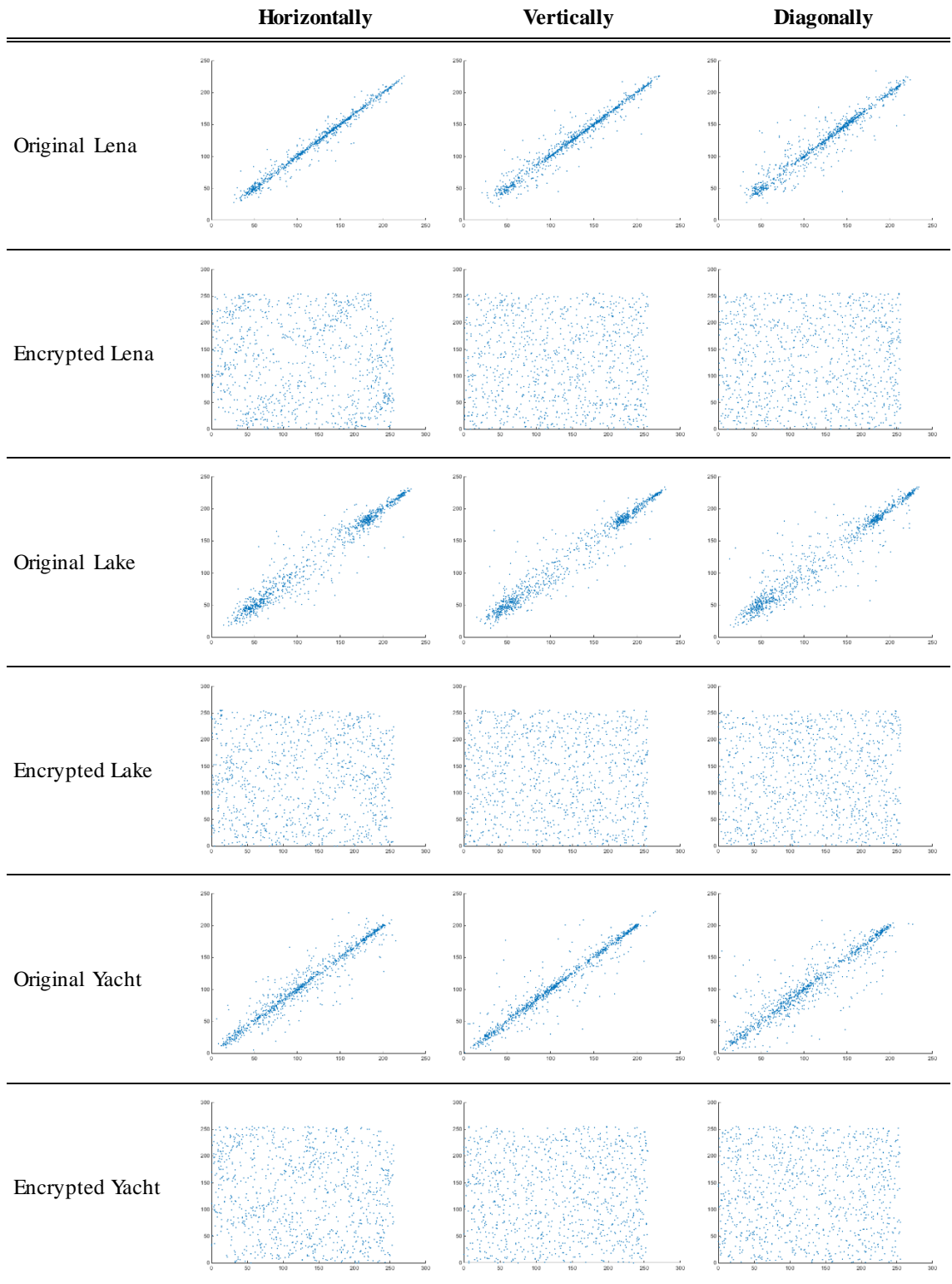

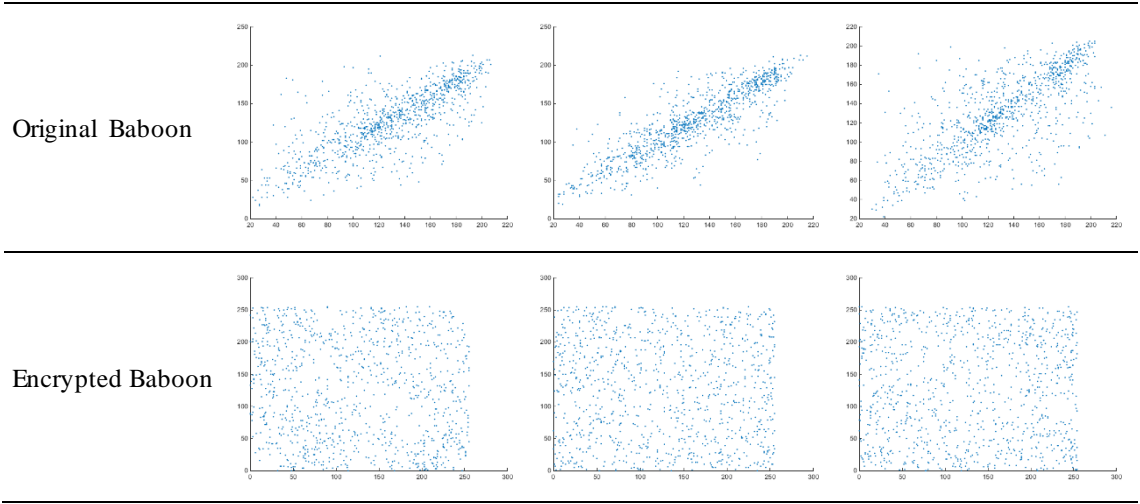

| | | |
|---|---|---|
| Original Baboon | | |
| Encrypted Baboon | | |

*Table 5.* Comparative study of NPCR and UACI of the proposed algorithm to some existing algorithms.

| Images | NPCR (%) | UACI (%) | Images | NPCR (%) | UACI (%) |
|---|---|---|---|---|---|
| Ref.[24](Lena) | 99.89 | 33.68 | Ref.[25](Lena) | 99.61 | 33.48 |
| Ref.[26](Lena) | 99.6972 | 33.5086 | Ref.[27](Lena) | 99.6086 | 33.4273 |
| Ref.[28](Lena) | 99.6000 | 33.5000 | Ref.[29](Lena) | 99.6100 | 33.4400 |
| Ref.[30](Lena) | 99.5983 | 33.6883 | Ref.[31](Lena) | 99.5865 | 33.4834 |
| Ref.[32](Lena) | 99.6399 | 33.5916 | Ref.[33](Lena) | 99.5926 | 33.3386 |
| Ours(Lena) | 99.4288 | 33.4320 | Ours(Lake) | 99.5966 | 33.5237 |
| Ours(Yacht) | 99.4586 | 33.5844 | Ours(Baboon) | 99.3977 | 33.3210 |

The above results prove that the proposed algorithm is feasible and robust for encrypting gray-level images. A color image can be divided into three channels: red channel, green channel and blue channel. Each channel can be viewed separately as a gray-level image. Which can be encrypted using our algorithm. The three encrypted channels are merged together to become an encrypted color image. Similarly, the encrypted color image can be decrypted in each channel, respectively. We have also used this algorithm to encrypt the color images: Lena, Lake, Airplane and Baboon, and it works really well. See the encryption effect in Table 6.

## 5 Conclusions

In the application of generating pseudo-stochastic sequential ciphers, the algorithm has the advantages of fast generation. By adjusting the number of neurons in the hidden layer, multiple sequences can be generated at a time. The experimental results show that the algorithm can be used for fast image encryption, and it has a very good significance for real-time remote image communication and image protection.

**Acknowledgements** This work was supported by Scientific and Technological Research Program of Chongqing Municipal Education Commission (No. KJ1501405, No. KJ1501409); Scientific and Technological Research Program of Chongqing University of Education (No. KY201522B, No. KY201520B); Fundamental Research Funds for the Central Universities (No. XDJK2016E068); Natural Science Foundation of China (No. 61170192) and National High-tech R&D Program (No. 2013AA013801).

Conflict of Interest: The authors declare that they have no conflict of interest.